%
\input phyzzx
\hfuzz 11pt
\font\mybb=msbm10 at 12pt

\def\Bbb#1{\hbox{\mybb#1}}

\def\bR{\Bbb {R}}
\def\bE{\Bbb {E}}

\def\bZ {\Bbb{Z}}

\def\dplus{=\hskip-5pt \raise 0.7pt\hbox{${}_\vert$} ^{\phantom 7}}
\def\dplusup{=\hskip-5.1pt \raise 5.4pt\hbox{${}_\vert$} ^{\phantom 7}}
\def\dplus{=\hskip-4.8pt \raise 0.7pt\hbox{${}_\vert$} ^{\phantom 7}}

\def\pmb#1{\setbox0=\hbox{#1} \kern-.025em\copy0\kern-\wd0
\kern0.05em\copy0\kern-\wd0 \kern-.025em\raise.0433em\box0}


\def\bfomega{\omega\kern-7.0pt \omega}

\REF\strom{A. Strominger, {\sl Open p-branes}, Phys. Lett. 
{\bf B383} 44; hep-th/9512059.}
\REF\pt{P.K. Townsend, {\sl Brane Surgery}, Nucl. 
Phys. Proc. Suppl {\bf 58} (1997) 163.}
\REF\gppkt{G. Papadopoulos \& P.K. Townsend,  
{\sl Intersecting M-branes}, Phys. Lett.
 {\bf B380} (1996) 273.}
\REF\malda{C.G. Callan, Jr and J.M. Maldacena, 
{\sl Brane dynamics from the
Born-Infeld action}, hep-th/9708147. }
\REF\gary{G.W. Gibbons, {\sl Born-Infeld particles 
and Dirichlet p-branes}, hep-th/9709027.}
\REF\towna{J. Gauntlett, J. Gomis \& P.K. Townsend, 
{\sl BPS bounds for worldvolume branes},
hep-th/9711205.}
\REF\kshir { K Shiraishi, {\sl Moduli space metric 
for maximally-charged dilaton black holes},
 Nucl.
Phys. {\bf B402} (1993) 399-410.}
\REF\kgg{G.W. Gibbons, G. Papadopoulos \& K.S. Stelle, {\sl HKT and OKT
geometries on soliton black hole moduli spaces},  
Nucl. Phys. {\bf B508} (1997) 623.}
\REF\coles{R. Coles \& G. Papadopoulos, {\sl The geometry 
of one-dimensional supersymmetric nonlinear
sigma models},  Class. Quantum Grav. {\bf 7} (1990) 427.}
\REF\gates{S.J. Gates, C.M. Hull \& M. Ro\v cek,  
{\sl Twisted multiplets and  new
supersymmetric nonlinear sigma models}, Nucl. Phys.  {\bf B248} (1984) 157.}
\REF\howepap{P.S. Howe \& G. Papadopoulos, {\sl Ultraviolet 
behaviour of two-dimensional 
supersymmetric nonlinear sigma models}, Nucl. Phys.
{\bf B289} (1987) 264; {\sl Further remarks on the geometry of 
two-dimensional nonlinear sigma models},
Class. Quantum Grav. {\bf 5} (1988) 1647;  {\sl Finiteness and 
anomalies in (4,0)-supersymmetric sigma
models}, Nucl .Phys.  {\bf B381} (1992) 360.}
\REF\manton{N. Manton, {\sl A remark on the scattering of BPS 
monopoles}, Phys. Lett. {\bf 110B} (1982)
54.}
\REF\gibrub {G W Gibbons \& P J Ruback, {\sl The Motion of extreme 
Reissner-Nordstr\"om black holes in
the low velocity limit},  Phys. Rev. Lett. {\bf 57}
 (1986) 1492.}
\REF\rub{ P J Ruback, {\sl The motion of Kaluza-Klein monopoles},  
Commun. Math. Phys. {\bf 107}
(1986) 93.}
\REF\gibkal{ G. W. Gibbons \& R. Kallosh,  {\sl Topology,
 entropy and the Witten index
of dilaton black holes}, Phys. Rev. {\bf
D51}
 (1995) 2839.}
\REF\ferear{ R C Ferrell \& D M Eardley, {\sl Slow motion scattering 
and coalescence of maximally
charged black holes},  Phys. Rev. Lett.  {\bf 59} (1987) 1617.}   
\REF\felsam{ A G Felce \& T M Samols, {\sl Low-energy dynamics of 
string solitons},  Phys. Letts.
{\bf B308}  (1993) 30: hep-th/921118.}
\REF\modfive{D.M. Kaplan and J. Michelson, {\sl Scattering of several
multiply charged extremal D=5 black holes}, hep-th/9707021.}
\REF\callan{C. G. Callan, J. Harvey \& A.  Strominger, 
{\sl Supersymmetric String Solitons},
hep-th/9112030.}
\REF\green{C. Bachas, M. Green \& A. Schwimmer, {\sl $(8,0)$ quantum
 mechanics and symmetry enhancement
in type I' superstrings}, hep-th/9712086.}
\REF\anders{M. Cederwall \& A. Westerberg, {\sl World-volume fields, 
$SL(2, \bZ)$ and duality: the type
IIB 3-brane}, hep-th/9710007.}
\REF\howe{P.S. Howe, N.D. Lambert \& P.C. West, {\sl The self-dual 
string soliton}, hep-th/9709014.}
\REF\ver{R. Dijkgraaf, E. Verlinde and H. Verlinde, {\sl BPS spectrum 
of the five-brane and black hole
entropy}, 
Nucl. Phys. {\bf B486} (1997) 77.}
\REF\nicolai{B. de Wit, A.K. Tollst\'en \& H. Nicolai, {\sl Locally 
supersymmetric $D=3$ non-linear
sigma models}, Nucl. Phys. {\bf B392} (1993) 3. }
\REF\gp{G. Papadopoulos, {\sl T-duality and the worldvolume solitons of 
five-branes and KK-monopoles },
hep-th/9712162.}
\REF\paulina{G.W. Gibbons, G. Papadopoulos 
and P. Ryckenkova, manuscript in preparation.}

\Pubnum{ \vbox{ \hbox{R/98/10}\hbox{} } }
\pubtype{}
\date{February, 1998}
\titlepage
\title{The Moduli Spaces of  Worldvolume Brane Solitons}
\author{J. Gutowski and G. Papadopoulos}
\address{DAMTP,\break Silver Street, \break University of 
Cambridge,\break Cambridge CB3 9EW}
\abstract { We compute the moduli metrics of worldvolume 0-brane solitons
of D-branes and the   worldvolume self-dual string solitons of the M-5-brane
 and examine their
geometry.  We find that the moduli spaces of 0-brane solitons of 
D-4-branes and D-8-branes  are 
hyper-K\"ahler  manifolds with torsion and
octonionic K\"ahler manifolds with torsion, respectively. The moduli 
space of the self-dual string
soliton of the M-5-brane is also  a hyper-K\"ahler manifold with torsion. }

\endpage
\pagenumber=2



\def\C{\mkern1mu\raise2.2pt\hbox{$\scriptscriptstyle|$}\mkern-7mu{\rm C}}



\chapter{ Introduction}

In the last few years much progress has been done to understand
 the various properties
of branes. It is remarkable that branes can intersect or end
 on other branes preserving
a proportion of the bulk supersymmetry [\strom, \pt, \gppkt]. 
From the perspective of  a
brane, these intersections and boundaries manifest themselves as 
classical supersymmetric solutions of
the associated worldvolume effective theory. Many such solutions 
have been found in [\malda, \gary];
they are commonly called worldvolume brane solitons and we shall
 use this convention\foot{We
remark though that some of worldvolume brane solitons are singular
 and do not interpolate between
different vacua of the effective theory.}. The  worldvolume solitons
 preserve the same proportion of
supersymmetry as  the associated bulk configurations and saturate a 
BPS bound [\towna]. Fundamental
strings end on D-branes. This leads to the presence of  worldvolume 
0-brane solitons  on all D-branes. 
These  have been found by G.W. Gibbons [\gary] as classical solutions 
of the Born-Infeld (BI) action.
The 0-brane soliton of a D-p-brane is
$$
\eqalign{
Y&=H
\cr
F_2&=dt\wedge dH\ ,}
\eqn\mone
$$
where $Y$ is a transverse scalar of the D-p-brane, $F_2$ is the 
two-form BI field strength\foot{Our
conventions for a p-form $\omega$ are $\omega={1\over p!} 
\omega_{i_1\dots i_p} dx^{i_1}\wedge\dots
\wedge dx^{i_p}$.} and 
$$
H=1+\sum_{i=1}^N {\mu_i\over |x-y_i|^{p-2}}\ ,
\eqn\mtwo
$$  
is a harmonic function of the  D-p-brane worldvolume coordinates 
$\{x^a; a=1,\dots,p\}$ transverse to
the 0-brane, {\sl i.e.} $\{x^a; a=1,\dots,p\}$ are the spatial 
worldvolume coordinates of the
D-p-brane. The centres
$\{y_i; i=1,\dots, k\}$ of the harmonic function are the positions
 of the 0-branes.
This solution preserves $1/4$ of the supersymmetry of
the bulk. The moduli space ${\cal M}^p_k$ of $k$-indistinguishable 
0-brane solitons of a D-p-brane is
the  configuration space $C_k(\bE^p)$
of $k$  particles propagating in $\bE^p$.

In this letter we shall present the moduli metric of the 0-brane 
solitons of D-branes. We shall find
that the moduli space geometry of the 0-branes is similar to
 that of the $a^2=1$ black-holes
of [\kshir, \kgg]. In particular, the moduli spaces of the 0-branes of a
D-4-brane and of a D-8-brane are hyper-K\"ahler manifolds with 
torsion (HKT) and octonionic-K\"ahler
manifolds with torsion (OKT), respectively. The effective 
theories of the 0-brane solitons are
one-dimensional sigma models with eight supersymmetries [\coles]. 
The sigma model associated with the HKT
geometry is the reduction of the two-dimensional (4,4)-supersymmetric
  one [\gates, \howepap] 
to one-dimension. The OKT geometry is particular to the target 
spaces of one-dimensional sigma models
with eight supersymmetries.  Then we shall describe the moduli
 spaces of other worldvolume solitons.
In particular we shall find that the moduli space of the 
self-dual string of the M-5-brane is an HKT
manifold. The associated effective theory is a 
two-dimensional (4,4)-supersymmetric sigma model with
Wess-Zumino term.

This letter is organized as follows: In section two, we 
give the moduli metrics of the
worldvolume 0-brane solitons of D-branes. In section three, we 
describe the effective actions of these
solitons.
In section four, we find the moduli metric of the self-dual 
string soliton of the
M-5-brane, and in section five we give our conclusions.

\chapter{The moduli metric of 0-branes}

The moduli metric of 0-brane solitons of a  D-p-brane is
 determined in a similar way  to that
of the the moduli metric of BPS monopoles [\manton] and  
solitonic black holes [\gibrub-\modfive]. The 
0-branes are solutions of the classical field equations of 
the BI actions which are the
effective theories of D-branes.  Therefore,
 the moduli metrics of the 0-branes are determined by the BI 
actions in the slow motion limit. However,
it suffices to consider the part of the BI action which is 
quadratic in the fields. We have also done
the calculation of the moduli metrics using the full BI 
action and  have obtained the same
results\foot{The details of this computation will be 
given elsewhere.}. Let $F_2$ be the BI two-form
field strength and
$\{Y^i; i=1,
\dots, 9-p\}$ be the transverse scalars of a D-p-brane 
in the static gauge. Expanding the BI action,
the term quadratic in the fields is
$$
S_{BI}={1\over2} \int d^{p+1}x \big(\delta_{ij}
\eta^{\mu\nu}\partial_\mu Y^i \partial_\nu
Y^j+{1\over2}F_{\mu\nu} F^{\mu\nu}\big)\ .
\eqn\bione
$$
The 0-branes, and the other worldvolume solitons that we 
shall investigate later, have
one non-vanishing transverse scalar $Y$. It turns out 
that for the computation
of the moduli metric,  we can
set
$Y=Y^1$ and allow
 the rest of the transverse scalars $\{Y^i; i=2, \dots, 9-p\}$ 
 to vanish. The resulting action is
$$
S_0={1\over2} \int d^{p+1}x \big(\eta^{\mu\nu}\partial_\mu Y \partial_\nu
Y+{1\over2}F_{\mu\nu} F^{\mu\nu}\big)\ .
\eqn\bionee
$$
We
remark that the 0-brane solution solves the field
 equations of \bionee. In addition the
action \bionee\ vanishes when it is evaluated at the 0-brane configuration.

 To compute
the moduli metric, we allow  the positions $\{y_i; i=1,\dots k\}$ 
of the 0-brane
solution to depend on
 the worldvolume time coordinate $t=x^0$ of the D-p-brane. Then we use 
the ansatz for the  BI field
$$
A_\mu dx^\mu= -H dt+B_a dx^a
\eqn\gauge
$$
and solve the field equations up to linear terms in the velocities,
where $\{\mu,\nu=0, \dots, p\}$ and $\{a, b=1,\dots,p\}$. The solution for $B$
is
$$
dB=-\sum_{i=1}^k \,\delta_{ab}
{\buildrel  . \over  y}_i^a dx^b\wedge dH_{(i)}  \ ,
\eqn\msol
$$
where ${\buildrel  . \over y}_i^a={d\over dt}y_i^a$ and 
$$
H_{(i)}={\mu_i\over |x-y_i|^{p-2}} \ .
\eqn\mthree
$$
To proceed, we  add   sources  to the action \bionee\ 
for both the BI field and
scalar field $Y$. These are
$$
S_{sourc}=(2-p) {\rm Vol}(S^{p-1})\sum^k_{i=1}\int\, dt\,  
\big(\mu_i Y+ \mu_i A_\mu {\partial
y^\mu_i\over
\partial s_i}\big)\ ,
\eqn\mfour
$$
where $\{s_i; i=1,\dots, k\}$ are the proper times of the 
0-branes and ${\rm Vol}(S^{p-1})$
is the volume of (p-1)-sphere of unit radius.
The total action is
$$
S=S_0+S_{sourc}\ .
\eqn\mfive
$$
Substituting $Y=Y(|x-y_i(t)|)$ and \gauge\ into this
 action using \msol, we find that
$$
S={p-2\over2}{\rm Vol}(S^{p-1})\, \int\, dt\, 
\big[\sum_{i=1}^k \mu_i |v_i|^2+ \sum^k_{i<j}\mu_i \mu_j {
|v_i-v_j|^2\over |y_i-y_j|^{p-2}}\big]\ ,
\eqn\mseven
$$
where $v_i^a={\buildrel  . \over y}_i^a$, {\sl i.e} $v_i$ is the
 velocity of the i-th 0-brane; all
the norms are taken with respect to the p-dimensional Euclidean metric.  
So the moduli metric is
$$
ds^2=\sum_{i=1}^k  \mu_i |dy_i|^2+  
\sum^k_{i<j} \mu_i \mu_j { |dy_i-dy_j|^2\over
|y_i-y_j|^{p-2}}\ .
\eqn\meight
$$

The metric is Galilean invariant and it is the sum of two parts. 
The first part of the metric
is due to the center of mass motion of the 0-brane solitons 
and the rest of the metric is due to the
relative motion. The center of mass motion decouples from the rest.  
The interactions of the
0-branes are two body interactions.

The  metric of the moduli space of the 0-brane  solitons
of a D-p-brane is the same as the metric of the moduli space of 
$a^2=1$ (p+1)-dimensional solitonic
black holes found in [\kshir]. In the string or M-theory picture, 
the worldvolume 0-branes and the
solitonic
$a^2=1$ black holes have common origin. Both are due to
the bulk ten-dimensional configuration of a fundamental string 
ending on the D-p-brane (up possibly to a
U-duality transformation). The black holes are found by reducing 
the associated supergravity solutions
to an appropriate dimension along the relative transverse 
directions of the configuration. While the
0-branes are the boundaries of the fundamental strings ending 
on the D-p-brane viewed from the
perspective of the worldvolume theory of the D-p-brane. 
Despite this, the moduli space of 0-branes and the moduli space 
of black holes describe different
motions. The moduli space derived from the BI action describes 
the motion of the 0-branes along the
D-p-brane and the moduli space of the black holes describes the 
motion of black holes  along the
overall transverse directions of the bulk configuration. 

\chapter{Supersymmetry and effective actions}

The 0-brane solitons preserve $1/4$ of the supersymmetry of the bulk. 
Therefore it is expected that
their effective theories are described by supersymmetric
 non-relativistic particle actions  with eight
supercharges. These are supersymmetric sigma models and  
have  been investigated
in [\coles].

An investigation of the supersymmetry projections  of a 
fundamental string ending on the D-4-brane
reveals that the effective theory of the 0-branes has
 (4,4) supersymmetry from the string perspective.
The  moduli metric of the 0-brane solitons of the D-4-brane is
$$
ds^2=\sum_{i=1}^k  \mu_i |dy_i|^2+  
\sum^k_{i<j} \mu_i \mu_j { |dy_i-dy_j|^2\over
|y_i-y_j|^{2}}\ .
\eqn\hktone
$$
It has been shown in [\kgg] that this metric admits two hyper-K\"ahler
 with torsion (HKT) structures.
The complex structures for the first HKT structure are 
$$
{\bf I}_r=I_r\otimes 1_k
\eqn\chone
$$
where $\{I_r; r=1,2,3\}$ are the constant complex structures on 
$\bE^4$ associated with the
self-dual 2-forms and $1_k$ is the identity $k\times k$ matrix. The complex 
structures of the 
other  HKT
structure are
$$
{\bf J}_r=J_r\otimes 1_k
\eqn\chtwo
$$
where $\{J_r; r=1,2,3\}$ are the constant complex 
structures on $\bE^4$ associated with the
anti-self-dual 2-forms on $\bE^4$.

The reduction of the (4,4)-supersymmetric two-dimensional 
sigma model action to one dimension
leads to an action for the  N=8a supersymmetry multiplet (in the
notation of [\kgg]). Given the moduli metric \hktone\ and the 
complex structures \chone\ and 
\chtwo, the rest of the effective action of the N=8a 
multiplet is determined. The
 complex structures  $I_r, J_r$ can be constructed from the 
killing spinors, as done in the
case of the heterotic 5-brane  where a similar geometry 
arises [\callan], and then induced on the moduli
space ${\cal M}_k$ as in \chone\ and \chtwo\ above. This 
is similar to the way that the
complex structures on the moduli spaces of BPS monopoles 
and instantons are induced
from the complex structures of the underlying space(time).

An investigation of the supersymmetry projections  of a 
fundamental string ending on the D-8-brane
reveals that the effective theory of the 0-branes has 
(8,0) supersymmetry from the string perspective. 
The moduli metric of the 0-brane solitons of the D-8-brane is
$$
ds^2=\sum_{i=1}^k  \mu_i |dy_i|^2+  
\sum^k_{i<j} \mu_i \mu_j { |dy_i-dy_j|^2\over
|y_i-y_j|^{6}}\ .
\eqn\oktone
$$
It has been shown in [\kgg] that this metric admits an 
octonionic-K\"ahler structure with torsion
(OKT). Therefore, the associated effective theory is
 based on the N=8b one-dimensional sigma model
supersymmetry multiplet.  The rest of the effective action 
including the fermion couplings depends on the choice of 
complex structures associated with the
N=8b multiplet. A similar argument to the one used for the 
previous case leads to a choice
of constant complex structures on $\bE^8$ associated with a 
basis in ${\rm Cliff}(\bE^8)$ equipped with
the negative definite inner product. Then the rest of the couplings 
of the effective theory are uniquely
determined from the metric
\oktone. However unlike the one-dimensional  N=8a 
supersymmetric multiplet, the N=8b one
does not have a two- or higher-dimensional origin.  It is worth 
mentioning that in [\green] (see also
references within) the motion of a D-0-brane along directions 
transverse to a D-8-brane has been
investigated leading again to a  one-dimensional 
effective theory with eight supercharges.

 The relation between the geometry of the moduli space of the 
0-brane solitons of D-p-branes for
$4<p<8$ and the supersymmetry of the associated 
effective action can be investigated as for $p=4,8$. It turns out 
that one can find supersymmetric
one-dimensional sigma model actions with eight supercharges and 
with a kinetic term for the bosons
determined by the moduli metric [\paulina]. 
 The effective action of the 0-brane solitons of the D-3-brane 
can be determined in a similar way.
However
$(p,q)$-strings ($(p,q)$ are coprime integers) can also end on 
the D-3-brane. The boundary then is a
0-brane which carries a (p,q)-dyon charge. Such 0-brane 
dyons should be solutions of the 
$SL(2,\bZ)$ invariant worldvolume action of the D-3-brane found in [\anders].

\chapter{The moduli metric of the self-dual string}

The solution of the self-dual string soliton of the M-5-brane
 has been given in [\howe]. Let $Y$ be a
transverse scalar of the M-5-brane and $F_3$ be a 
self-dual 3-form. Then the solution is
$$
\eqalign{
Y&=H
\cr
F_3&= f_3+*f_3}
\eqn\mnine
$$
where
$$
f_3={1\over4}dt\wedge d\rho\wedge dH\ ,
\eqn\mten
$$
$\rho$ is the spatial worldvolume self-dual string direction 
and $H$ is a harmonic function
in $\bE^4$ spanned by the worldvolume directions of the 
M-5-brane transverse to the self-dual string.
The Hodge dual is taken by the flat metric on $\bE^{(1,5)}$.
This solution is invariant under the action of the 
Poincar\'e group on the worldvolume
coordinates $(t,\rho)$ of the self-dual string soliton. In particular
$\xi=\partial/\partial \rho$ leaves the solution invariant.

From the bulk perspective, the self-dual string is the 
boundary of a M-2-brane ending on 
an M-5-brane.  Compactifying this configuration 
along the self-dual string worldvolume
direction $\rho$ to ten-dimensions leads to a configuration with 
the interpretation of a IIA fundamental
string ending on a D-4-brane. The M-5-brane field 
equations wrapped along a worldvolume
direction reduce to those of the D-4-brane. Moreover we 
have verified that the self-dual
string soliton of the M-5-brane reduces along the 
direction $\rho$ to that of the 0-brane  of
the D-4-brane. Therefore we conclude from this that 
the metric on the moduli space
of self-dual string solitons is the same as that of the 
moduli space of 0-brane solitons
of D-4-branes found in section 2.

The effective theory of the self-dual string is described 
by a two-dimensional
(4,4)-supersymmetric sigma model. One can  easily see this 
by  examining the supersymmetry projections
of the associated bulk configuration. The same point has 
also been argued in [\ver].
The effective action of the self-dual string is determined 
by lifting to two dimensions the
effective action of the  0-branes of the D-4-brane. This is 
consistent because, as we have mentioned in
the previous section,  the N=8a one-dimensional supersymmetry 
multiplet  is the reduction of the
two-dimensional  (4,4)-supersymmetric one. Choosing the complex
structures to be constant as in the 0-brane effective action, 
we find that the effective
action of the self-dual string, apart from the kinetic term 
associated with the metric \hktone,
  has a Wess-Zumino term $b$.  This term is  given by
$$
c=-{1\over9}i_I d\omega_I
\eqn\meleven
$$
where $c=db$, $I$ is one of the complex structures and $\omega_I$ is 
the associated K\"ahler form
with respect to the metric \hktone. This term is not present in 
the effective action of the 0-branes
because it vanishes upon reduction to one dimension. We remark 
though that the fermion couplings of the
one-dimensional effective action depend on
$c$ [\coles]. The coefficient of the Wess-Zumino term is 
quantized. To see this in the case of 
the moduli space of two self-dual string solitons, we 
observe that the topology of the space is
${\cal M}_2^4=\bR^4\times \bR^+\times S^3$ and that the 
Wess-Zumino term in the volume of $S^3$.

\chapter{Concluding Remarks}

We have found the moduli metrics of the 0-brane 
solitons of the D-p-branes and the
self-dual string soliton of the M-5-brane.  In 
particular the
moduli space of the 0-brane solitons of the D-4-brane 
and the moduli space of the self-dual string
soliton
 of the M-5-brane are HKT manifolds. Then we have
investigated their effective actions and we have found 
that they are described by one-and
two-dimensional supersymmetric sigma models with eight supersymmetries.
Fixing the complex structures associated with the 
sigma model supersymmetry multiplets,
the effective actions are determined uniquely from the moduli metric.

There are other worldvolume  solutions such as  the 2-branes solitons of the
IIB NS-5-brane and the IIB D-5-brane. The moduli spaces of
 these two solitons should be the
same because they are related by S-duality. The associated 
effective theory of these two solitons is
a 3-dimensional sigma model with eight supercharges (after 
possibly dualizing vector multiplets to
scalar ones).  Then supersymmetry requires that the target
 space of the sigma model, and so the moduli
space, is a hyper-K\"ahler manifold [\nicolai]. Another 
worldvolume solution is the 0-brane soliton of
the IIB NS-5-brane. This is related to the 0-brane soliton 
of the IIB D-5-brane by S-duality. Therefore
it is expected that the moduli space of the 0-brane soliton 
of IIB NS-5-brane is the same as that
of the 0-brane of the IIB D-5-brane. 

Another class of worldvolume solitons is that of  the KK-monopoles [\gp].
However, these solitons are related by T-duality to those 
of the IIA and IIB NS-5-branes,
and it is expected that their moduli spaces can be 
determined from those of the worldvolume
solitons of the IIA and IIB NS-5-branes.
\vskip 1cm
\noindent{\bf Acknowledgments:}   We thank G.W. Gibbons and N. Manton for
 helpful discussions.  G.P.
is supported by a University Research Fellowship from the Royal
Society. J.G thanks EPSRC for a studentship.

\refout

\bye